\documentclass[twocolumn,showpacs,preprintnumbers,amsmath,amssymb,pre]{revtex4}

\usepackage{graphicx,amsfonts}

\begin{document}

\title{Slip behavior in liquid films on surfaces of patterned wettability:\\ 
Comparison between continuum and molecular dynamics simulations}

\author{Nikolai V. Priezjev, Anton A. Darhuber, and Sandra M. Troian}
\email{stroian@princeton.edu} \homepage{http://www.princeton.edu/~stroian}

\affiliation{Microfluidic Research \& Engineering Laboratory, School of Engineering \&
Applied Science, Princeton University, Princeton, New Jersey 08544}

\date{\today}

\begin{abstract}
We investigate the behavior of the slip length in Newtonian liquids subject to planar
shear bounded by substrates with mixed boundary conditions. The upper wall, consisting of
a homogenous surface of finite or vanishing slip, moves at a constant speed parallel to a
lower stationary wall, whose surface is patterned with an array of stripes representing
alternating regions of no-shear and finite or no-slip. Velocity fields and effective slip
lengths are computed both from molecular dynamics (MD) simulations and solution of the
Stokes equation for flow configurations either parallel or perpendicular to the stripes.
Excellent agreement between the hydrodynamic and MD results is obtained when the
normalized width of the slip regions, $a/\sigma \gtrsim {\cal O}(10)$, where $\sigma$ is
the (fluid) molecular diameter characterizing the Lennard-Jones interaction. In this
regime, the effective slip length increases monotonically with $a/\sigma$ to a saturation
value. For $a/\sigma \lesssim {\cal O}(10)$ and transverse flow configurations, the
non-uniform interaction potential at the lower wall constitutes a rough surface whose
molecular scale corrugations strongly reduce the effective slip length below the
hydrodynamic results. The translational symmetry for longitudinal flow eliminates the
influence of molecular scale roughness; however, the reduced molecular ordering above the
wetting regions of finite slip for small values of $a/\sigma$ increases the value of the
effective slip length far above the hydrodynamic predictions. The strong
correlation between the effective slip length and the liquid structure factor
representative of the first fluid layer near the patterned wall illustrates the influence
of molecular ordering effects on slip in non-inertial flows.
\end{abstract}

\maketitle

\section{Introduction} \label{sec:Introduction}

The development of micro- and nanofluidic devices for the manipulation of films, drops
and bubbles requires detailed knowledge of interfacial phenomena and small scale flows.
These systems, which are distinguished by a large surface-to-volume ratio and flow at
small Reynolds, capillary and Bond numbers, are strongly influenced by boundary effects
\cite{Darhuber05}. Liquid affinity to nearby solid boundaries can be reduced through
chemical treatments \cite{Schnell56,Baudry01,Meinhart02,Breuer03}, substrate topology
\cite{Watanabe99,Bocquet03,Krupenkin} or the nucleation of nanobubbles on hydrophobic
glass surfaces \cite{Ishida00,Tyrrell01,Steitz03}. Weak van der Waals interactions between
a polymer melt and solid wall \cite{Leger97,Denn01,Leger03} or between two immiscible
polymers \cite{Wyart90} can also lead to significant slippage and reduced frictional
resistance. The degree of slip is normally quantified through the slip length defined as
the distance from the surface within the solid phase where the extrapolated flow velocity
vanishes \cite{Navier}. Numerous experimental and theoretical studies have examined how
the slip length is influenced by such factors as the degree of hydrophobicity
\cite{Schnell56,Vinogradova99}, the substrate topography and surface roughness
\cite{Watanabe99,Bocquet03,Richardson73,Hocking76,Jansons88,Einzel90,Davis94,Barrat94,Cottin},
the presence of interstitial lubricating layers \cite{Davis94,Barnes95,Vinogradova03},
the polymer molecular weight \cite{Migler93,Archer,Denn01} and the applied shear rate
\cite{Nature97,Priezjev04,Vinogradova00,Granick01,Breuer03}. In a recent development, the
large values of the slip length extracted from experiments involving the pressure-driven
flow of water through hydrophobically coated capillaries have been attributed
\cite{Vinogradova95,Stone03} to the spontaneous nucleation of a dense and stable layer of
nanobubbles in water films adjacent to hydrophobic glass surfaces
\cite{Ishida00,Tyrrell01,Steitz03}. Of special interest is the corresponding reduction in
drag achieved by proportional substitution of liquid-solid contact area with liquid-gas
contact area or equivalently, substitution of regions of no-slip or finite slip by
regions of essentially no-shear (i.e. infinite slip).

Interest in the hydrodynamic behavior of liquid films in the vicinity of surfaces with
mixed boundary conditions dates back several decades to the work of Philip
\cite{Philip1,Philip2}. He examined the steady flow of an incompressible and inertia-less
Newtonian liquid driven either by a uniform shear stress or uniform pressure gradient and
subject to mixed wall boundary conditions. These were represented by surfaces consisting
of alternating striped regions of no-shear and no-slip among other geometries. Using
conformal mapping, Philip \cite{Philip1,Philip2} derived analytic expressions for the
streamfunction and volumetric flux for flow perpendicular (transverse configuration) or
parallel (longitudinal configuration) to the striped array in the limit of Stokes flow. Recently,
Lauga and Stone~\cite{Stone03} investigated the behavior of the effective slip length for
steady Poiseuille flow through a capillary of circular cross-section whose inner wall
consists of periodically distributed regions of no-slip and no-shear. Philip's earlier
treatment was used to extract the slip length for longitudinal configurations; additional
analysis was required for transverse configurations. Comparison of their results with
available experimental measurements suggests what model parameter values would reproduce
the experimental slip lengths. For slip lengths in the nanometer range, one might ask
whether a hydrodynamic analysis can correctly predict these values or whether the
molecular aspects of the fluid can strong influence the slip behavior causing deviations
from the continuum theory.

\begin{figure}[t]
\begin{center}
\includegraphics[width=7.0cm]{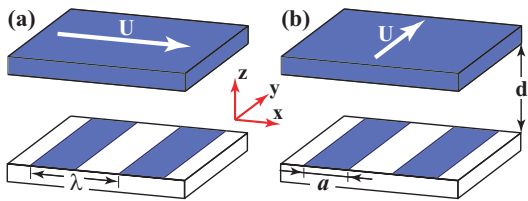}
\end{center}
\caption{(Color online) (a) Transverse and (b) longitudinal flow orientations for a
liquid film subject to planar shear in a cell with wall separation $d$. Darker stripes of
width $a$ signify regions of finite slip or no-slip. White stripes signify regions of
no-shear (or equivalently perfect slip). The upper wall moves at constant speed $U$
relative to the lower stationary surface ($z\!=\!0$). The periodicity of the lower wall
pattern geometry is designed by $\lambda$.} \label{Couette_geometry}
\end{figure}

Molecular dynamics (MD) simulations provide an ideal tool for investigating the
conformation and behavior of fluid molecules adjacent to chemically or topologically
textured substrates. The boundary conditions which establish the flow profile are not
specified {\it a priori}, but arise naturally from the wall-fluid contrast in density and
the fluid-fluid and wall-fluid interaction potentials. In recent years, many groups 
have examined how various
molecular parameters characterizing the wall and fluid properties affect the degree of
slip at liquid-solid interfaces. In particular, it has been demonstrated that the
structure factor and contact density representative of the first fluid layer adjacent to
a wall significantly influence the degree of slip in Newtonian and non-Newtonian fluids
\cite{Thompson90,Thompson92,Thompson95,Barrat94,Nature97,Barrat99,Priezjev04}. The
results of this current study confirm the importance of these molecular parameters
for flow on heterogeneous substrates.

In this work, we investigate the behavior of the slip length in viscous films under
planar shear bounded by substrates with mixed boundary conditions using both molecular
dynamics (MD) simulations and Stokes flow computations. The upper wall, consisting of a
homogenous surface of finite or no-slip, moves at a constant speed, $U$,  a distance $d$
above a lower stationary wall, whose surface is patterned with an infinite array of
stripes representing alternating regions of no-shear and finite or no-slip. As shown in
Fig.~\ref{Couette_geometry}, we consider transverse and longitudinal flow configurations
and compute the corresponding velocity fields and effective slip lengths for a wide range
of stripe widths, periods and liquid-solid affinities. Excellent agreement between the
hydrodynamic and MD results is obtained when the normalized width of the slip regions,
$a/\sigma \gtrsim {\cal O}(10)$, where $\sigma$ is the (fluid) molecular diameter
characterizing the Lennard--Jones interaction. For surface patterns approaching molecular
size, the degree of fluid ordering near the patterned wall, as quantified by the in-plane
structure factor and contact density in the first liquid layer, plays a dominant role
causing significant deviations from the hydrodynamic predictions. These deviations can be
explained in the context of effective surface roughness and molecular ordering
effects.

\section{Hydrodynamic Analysis} \label{sec:Continuum}

In the limit of vanishingly small Reynolds number ${\sf Re}\!=\!\rho U d/\mu$, where
$\rho$ and $\mu$ denote the (constant) liquid density and viscosity, inertial effects are
negligible. The velocity profile is then governed by the Stokes equation, $\nabla^2 {\bf
u} \!=\!\nabla p/\mu$, where the velocity field, ${\bf u}$, satisfies the condition of
incompressibility, $\nabla\!\cdot\!{\bf u}\!=\!0$, and $p$ denotes the pressure
distribution which in this study is induced by the patterned substrates. Application of
the divergence operator to the Stokes equation shows that the pressure field satisfies
the equation $\nabla^2 p\!=\!0$. It then follows that the velocity field satisfies the
biharmonic equation $\nabla^2 \nabla^2 {\bf u}=0$ \cite{Leal92}.

In the next section, we derive the boundary conditions (BCs) corresponding to transverse
[Fig.~\ref{Couette_geometry}(a)] and longitudinal [Fig.~\ref{Couette_geometry}(b)] flow
orientations. These conditions are used to compute numerical solutions of the
streamfunction, velocity field and effective slip length as a function of the
dimensionless stripe width of the finite slip regions, $a/\lambda$, and the dimensionless
surface period $\lambda/d$. The $\hat{y}$-axis is oriented parallel to the stripe edges
for either configuration. All numerical calculations were performed with the finite
element software {\sf FemLab}~2.3 \cite{Femlab,Wass04} using triangular elements with
quadratic basis functions. The solutions reported converged upon mesh refinement.

\subsection{Transverse configuration \label{Couette_trans}}

The two dimensional velocity field corresponding to the transverse configuration shown in
Fig.~\ref{Couette_geometry}(a) is represented by ${\bf u}(x,z)\!=\!(u,0,w)=\! (\partial
\psi/\partial z, 0, -\partial \psi/\partial x)$, where $\psi(x,z)$ denotes the
streamfunction, which implicitly satisfies the continuity equation $\nabla\!\cdot\!{\bf
u}\!=\!0$. The vorticity vector ${\bf \Omega}\!=\!\nabla \!\times {\bf
u}\!=\!(0,\omega,0)$, where $\omega=\partial u/\partial z\!-\!\partial w/\partial x$, has
only one non-zero component. According to these definitions, it follows that
\begin{equation}
\omega =\frac{\partial ^{2}\psi }{\partial z^{2}}+\frac{\partial ^{2}\psi }{\partial
x^{2}}={\nabla}^2 \psi\ \ \ \ \ \mbox{and}\ \ \ \ \ {\nabla}^2 \omega =0.
\label{vorticity_diffusion}
\end{equation}

\subsubsection{Boundary Conditions} \label{FullBCs}

\begin{figure}[tb]
\begin{center}
\includegraphics[width=8.3cm]{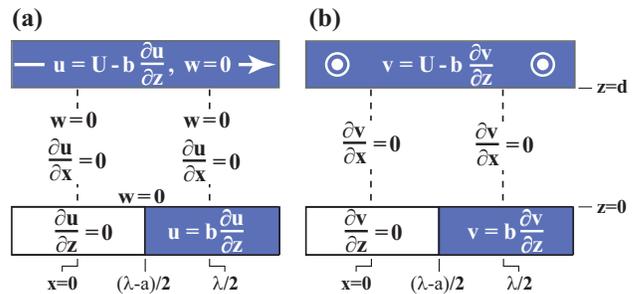}
\end{center}
\caption{(Color online) Computational domain and boundary conditions used for solution of
the Stokes equation corresponding to the (a) transverse and (b) longitudinal flow
orientation shown in Fig.~\ref{Couette_geometry}. The computational domain consists of
the region bounded by the upper and lower walls ($z\!=\!0$ and $z\!=\!d$) and the lateral
dashed lines ($x\!=\!0$ and $x\!=\!\lambda/2$), which are positioned at neighboring
midplanes of the no-shear and finite-slip regions.} \label{Fig_BCs}
\end{figure}

Solutions of the equations for the vorticity and streamfunction given by
Eq.~(\ref{vorticity_diffusion}) require the specification of eight BCs. The computational
domain sketched in Fig.~\ref{Fig_BCs}(a) is defined by the region bounded by the upper
and lower walls ($0\!\leq \!z\! \leq \!d$) and the dashed lines ($0\! \leq \!x\! \leq
\!\lambda/2$) corresponding to the midplanes of neighboring stripes. White surfaces
designate shear-free boundaries (i.e. surfaces of perfect slip); dark surfaces designate
boundaries of finite or no-slip. Throughout, partial derivatives are denoted by letter
subscripts e.g. $\partial \psi/\partial x \equiv \psi_x$.

The top and bottom walls represent impenetrable surfaces where
$w(x,z\!=\!0)=w(x,z\!=\!d)\!=\!0$, or in terms of the streamfunction, $\psi_x
(x,z\!=\!0)\!=\!\psi_x (x,z\!=\!d)\!=\!0$. The tangential component of the velocity field
must satisfy mixed slip and shear conditions at the upper and lower walls of the cell.
The no-shear BC is given by $u_z (0\! \leq \!x\! \leq \! \frac{\lambda - a}{2},
z\!=\!0)=0$. Slip surfaces are characterized by the Navier \cite{Navier} slip condition
$u\left[\frac{\lambda - a}{2}\! \leq \!x\! \leq \!\frac{\lambda}{2}, z\!=\!0\right] = b
u_z$ and $u[x, z\!=\!d]\!=\!U\!-\!b u_z$. The Navier slip length $b$ is assumed constant,
i.e. independent of the shear rate $\dot{\gamma}$.

The lateral boundary conditions for the {\em scalar} field $u$ are derived from the
following symmetry considerations. The biharmonic equation $\nabla^2 \nabla^2 u\!=\!0$
involves $x$-derivatives of even order only. The lower wall comprises an infinite number 
of mirror symmetry planes located at the at the stripe centers
$x\!=\!n\lambda/2$, for all integers $n$. Since the upper surface is homogeneous and
translationally  invariant, the mirror symmetry imposed by the lower surface determines
which symmetry applies throughout the entire Couette cell. The scalar field, $u$,
therefore also assumes mirror symmetry about the stripe centers such that
$u(x,z)\!=\!u(-x,z)$ and $u_x(x,z)\!=\! 0$ for all $x\!=\!n\lambda/2$ and integers $n$.
From the continuity equation, it then also follows that $w_z (x\!=\!n\lambda/2)=0$, i.e. $w$ is
independent of the coordinate $z$ within any mirror plane. Since the upper and lowers
walls are impenetrable, i.e. $w(x,z\!=\!0)=w(x,z\!=\!d)\!=\!0$, this constraint reduces
to the BC $\psi_x(x\!=\!0)=0=\psi_x(x\!=\!\lambda/2)$.

The continuity equation requires $u_x + w_z=0$. Together with the condition
$u(x,z)=u(-x,z)$, this implies $w(x,z)=-w(-x,z)$ such that $w(x,z) \!=\! 0$ 
and $w_{xx}(x,z)=0$ at all $x=n\lambda/2$ where $n=0, 1, 2 \ldots$. 
Substitution of this last relation and $u_x(x\!=\!0,z)=0$ into the expression
for $\omega_x$ leads to  $\omega_x(x\!=\!n\lambda/2,z)=0$. Regions of no-shear at the
lower wall are represented by the condition $\omega(x,z\!=\!0)= 0$.

Along the top and bottom walls, the scalar component $w$ is independent of the coordinate
$x$ and therefore $w_x(z\!=\!0)=w_x(z\!=\!d)=0$. Consequently, the vorticity at the top
and bottom walls reduces to $\omega=u_z$ and the Navier slip conditions can be rewritten
as $\psi_z\left(\frac{\lambda - a}{2}\! \leq \!x\! \leq
\!\frac{\lambda}{2},z\!=\!0\right) = b\omega$ and $\psi_z (x,z\!=\!d) = U-b\omega$. The
relation $w(x,z\!=\!0)=w(x,z\!=\!d)=\psi_x=0$ also implies that the streamfunction is
constant in the planes $z\!=\!0$ and $z\!=\!d$, whose values we denote by $\psi_{\rm
top}$ and $\psi_{\rm bottom}$. The difference in the streamfunction value between the top
and bottom walls is equal to the volumetric flux per unit length along the $y$-axis
\cite{Leal92} since
\begin{equation}
Q= \int \limits_{z=0}^{z=d}\! u(x,z)\,dz = \int \limits_{z=0}^{z=d}\! \psi_z \,dz =
\psi_{\rm top}- \psi_{\rm bottom}~.
\end{equation}
Because the streamfunction can only be determined within an arbitrary constant, we set
the value of $\psi_{\rm bottom}$ in these studies to zero without loss in generality. The
complete set of BCs for the vorticity and streamfunctions are therefore given by:
\begin{eqnarray}
\psi(x, z\!=\!0)   &=& 0 \label{BC1} \\
\psi(x,z\!=\!d)    &=& \psi_{\rm top} \\
\omega(x,z\!=\!0)  &=& 0\ \ \ \ \ \ \mbox{for $0\le x< \frac{\lambda -a}{2}$}\\
\psi_z(x,z\!=\!0)  &=& b\omega\ \ \ \ \mbox{for $\frac{\lambda-a}{2} \le x \le \frac{\lambda}{2}$}\\
\psi_z(x,z=d)      &=& U-b \omega\\
\psi_x(x\!=\!0,z)  &=& 0\ =\   \psi_x(x\!=\! \lambda/2,z)\\
\omega_x(x\!=\!0,z)&=& 0\ =\ \omega_x(x\!=\! \lambda/2,z)~. \label{BC6}
\end{eqnarray}

\subsubsection{Solution procedure} \label{Solnproc}

The value of $\psi_{\rm top}$ is determined from the pressure field as follows. The
Stokes equation for the vertical component of the velocity field is given by
$w_{xx}+w_{zz}\!=\! p_z/\mu$. As argued in a previous section, however,
$w_{xx}(x\!=\!n\lambda/2,z)\!=\!0$, and since $w$ is independent of $z$ along any mirror
symmetry plane, $w_{zz}(x\!=\!n\lambda/2,z)\!=\!0$. The pressure is therefore independent
of the vertical coordinate $z$ in all planes $x\!=\!n\lambda/2$. Furthermore, in the
absence of any \textit{externally} applied pressure gradient, as is the case here, and
because of the flow periodicity, $p(x\!=\!0)=p(x\!=\!\lambda)$. Since it was previously
argued that $u$ exhibits mirror symmetry about the planes $x\!=\!n\lambda/2$, it must
also be true of $p_x$ since $u_{xx}+u_{zz}\!=\! p_x/\mu$. Consequently, the pressure is
equal at the lateral boundaries of the computational cell, i.e. $p(x\!=\!0) =
p(x\!=\!\lambda/2)$. For convenience we set $p(x\!=\!0)\!=\!0$. This constraint, coupled
with the relation $p_x/\mu \!=\! -(\nabla \times {\bf \Omega})\cdot \hat{e}_x =
\omega_z$, was used to adjust the numerical value of $\psi_{\rm top}$ by requiring that
the following integral vanish identically:
\begin{equation}
\int_0^{\lambda/2}\!\! p_x\,dx = \mu\int_0^{\lambda/2}\! \omega_z\,dx\ =0~.
\end{equation}
The value of the effective slip length, $L_s$, corresponding to the overall flow within a
patterned cell was obtained from linear extrapolation of the averaged velocity profile
$\langle u \rangle\!=\!(2/\lambda)\int_0^{\lambda/2} u(x,z)\, dx$ to zero. Since at
planes of mirror symmetry, $u_x\!=\!0$ and $p(x\!=\!0)=p(x\!=\!\lambda/2)$, the integral
$\mu \int_0^{\lambda/2} \nabla^2 u(x,z)dx = \int_0^{\lambda/2} p_x dx\!=\!0$, reduces to
$\langle u \rangle_{zz}\!=\!0$. The averaged velocity field, $\langle u \rangle$, is
therefore a linear function of $z$ and geometric similarity establishes the relation for
the effective slip length, namely
\begin{equation} \label{slipdefine}
\frac{L_s}{d} = \frac{\langle u(z\!=\!0) \rangle}{\langle u(z\!=\!d) \rangle - \langle
u(z\!=\!0) \rangle}~~.
\end{equation}
For the numerical analysis, the equations for the vorticity and streamfunction given by
Eq.~(\ref{vorticity_diffusion}) and the BCs given by Eqs.~(\ref{BC1}-\ref{BC6}) were
non-dimensionalized according to the rescaled variables
\begin{eqnarray}\label{nondim3}
\widetilde{x}&=&x/\lambda\ \ \ \ \ \ \ \ \ \ \ \ \ \ \widetilde{z}=z/d\label{nondim1}\\
\widetilde{u}&=&u/U\ \ \ \ \ \ \ \ \ \ \ \ \ \widetilde{w}=w\left/\left( U \frac{d}{\lambda } \right)\right. \\
\widetilde{\psi} &=& \psi\left/\left(\frac{U d}{2}\right)\right.\ \ \ \widetilde{\omega
}=\omega \left/\left( \frac{U}{d}\right)\right.~,
\end{eqnarray}
leading to
\begin{equation} \label{final_streamf_eq1}
\frac{d^2}{\lambda^2}\frac{\partial^2 \widetilde{\psi}}{\partial \widetilde{x}^2} +
\frac{\partial^2 \widetilde{\psi} }{\partial\widetilde{z}^2} = 2 \widetilde{\omega}\ \ \
\ \ \mbox{and} \ \ \ \ \ \frac{d^2}{\lambda^2}\frac{\partial^2 \widetilde{\omega
}}{\partial \widetilde{x}^2}+\frac{\partial^2 \widetilde{\omega}}{\partial
\widetilde{z}^2}=0~.
\end{equation}
In Section \ref{sec:limits} we present numerical solutions to
Eqs.~(\ref{final_streamf_eq1}) and the extracted values of $L_s$ as a function of the
local slip length $b$ and pattern geometry. Analytic expressions are derived in the
limits $\lambda/d \rightarrow 0$ and $\lambda/d \rightarrow \infty$.

\subsubsection{Perturbative analysis for $b=0$}

In order to enhance the numerical precision of solutions corresponding to small values of
$L_s$, the velocity and pressure fields were decomposed into two contributions, ${\bf
u}\!=\!{\bf u}_0 + {\bf u}_1$ and $p\!=\!p_0 + p_1$. Here, ${\bf u}_0 \!=\! (Uz/d,0,0)$
and $p_0\!=\!0$ correspond to the velocity and pressure fields for planar shear flow
subject to no-slip at both solid boundaries. The Stokes equation then reduces to the form
$\mu \nabla^2 {\bf u}_1\!=\! \nabla p_1$, where the perturbed velocity field satisfies
the continuity equation $\nabla\! \cdot\! {\bf u}_1\!=\!0$. The following BCs for the
perturbed streamfunction and vorticity fields were determined in similar fashion as those
in Section~\ref{FullBCs}:
\begin{eqnarray}
\psi_1(x,z\!=\!0)     &=& 0\\
\psi_1(x,z\!=\!d)     &=& \psi_{1,{\rm top}}\\
\psi_{1,z}(x,z=d)     &=& U\\
\omega_1\left(\mbox{$0\!\le\! x \!<\! \frac{\lambda -a}{2}$},z\!=\!0\right)   &=& -U/d\\
\psi_{1,z}\left(\mbox{$\frac{\lambda-a}{2} \!\le\! x \!\le\! \frac{\lambda}{2}$},z\!=\!0\right) &=& 0\\
\psi_{1,x}(x\!=\!0,z)&=& 0\ =\   \psi_{1,x}\left(\mbox{$x\!=\!\frac{\lambda}{2}$},z\right) \\
\omega_{1,x}(x\!=\!0,z)&=& 0\ =\
\omega_{1,x}\left(\mbox{$x\!=\!\frac{\lambda}{2}$},z\right)
\end{eqnarray}
where $\psi_{1,z}\!=\!u_1$, $\psi_{1,x}\!=\!-w_1$ and $\omega_1\!=\!u_{1,z}-w_{1,x}$.
Non-dimensionalization of the vorticity and streamfunction perturbations $\omega_1$ and
$\psi_1$ as in Section~\ref{Solnproc} leads to:
\begin{eqnarray}
\frac{d^2}{\lambda^2}\frac{\partial^2 \widetilde{\psi}_1}{\partial \widetilde{x}^2} +
\frac{\partial^2 \widetilde{\psi}_1 }{\partial\widetilde{z}^2} &=& 2 \widetilde{\omega}_1\\
\frac{d^2}{\lambda^2}\frac{\partial^2 \widetilde{\omega}_1}{\partial
\widetilde{x}^2}+\frac{\partial^2 \widetilde{\omega}_1}{\partial \widetilde{z}^2}&=&0~.
\end{eqnarray}

\subsubsection{Numerical results and limiting cases} \label{sec:limits}

\begin{figure}[t]
\begin{center}
\includegraphics[width=7.6cm]{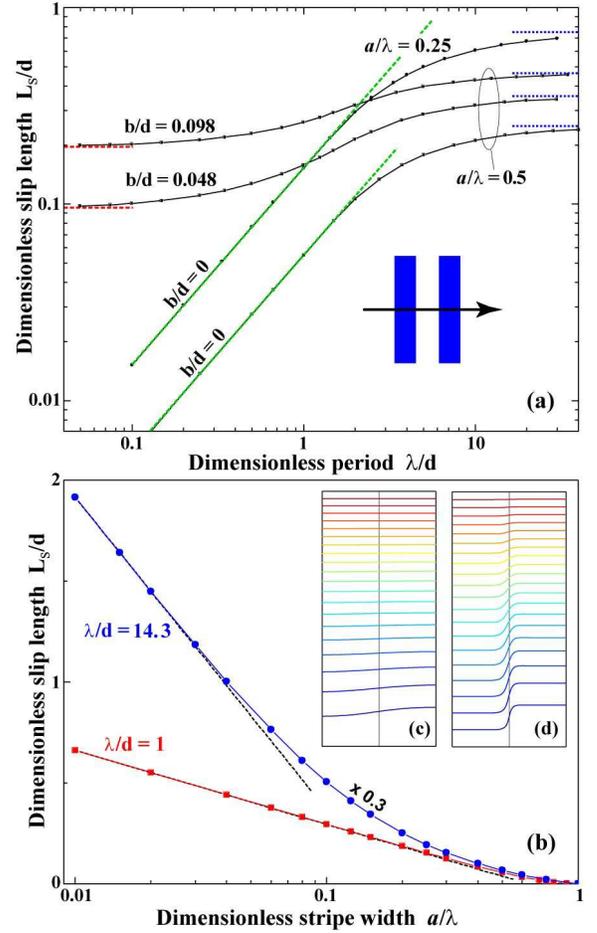}
\end{center}
\caption{(Color online) (a) Normalized slip length, $L_s/d$, versus normalized pattern
period, $\lambda/d$, derived from the Stokes solutions for the transverse flow
orientation. The parameters values shown are $b/d\!=\!0$, 0.048 and 0.098 and
$a/\lambda\!=\!0.25$ and 0.50. The straight lines superimposed on the numerical solutions
for $b/d\!=\!0$ correspond to the analytic limit $L_s/d \sim \lambda/d$.
(b) Normalized slip length $L_s/d$ versus normalized stripe width $a/\lambda$ for
$\lambda/d\!=\!1$ and 14.3 in the limit $b/d\!=\!0$. The dashed lines correspond to the
function $L_s/d = A\ln[a/\lambda]+B$, with fitting parameters $A$ and $B$. The data
points for the case $\lambda/d\!=\!14.3$ ($A\!=\!-2.245$, $B\!=\!-3.952$) are scaled by a
factor 0.3 for convenience.
(c-d) Streamlines corresponding to the transverse Stokes flow solutions for (c)
$\lambda/d\!=\!1$ and (d) $\lambda/d\!=\!20$ where $b/d\!=\!0.048$ and
$a/\lambda\!=\!0.5$.
The cell domain size is $\lambda/2 \times d$; the thin vertical lines designate the
boundary between surfaces of no-shear (left) and finite slip (right).}
\label{Couette_across_results}
\end{figure}

In Fig.~\ref{Couette_across_results}(a) is plotted the numerical results for the
normalized effective slip length, $L_s/d$, as a function of the aspect ratio,
$\lambda/d$, for the transverse configuration. Over the range shown, $L_s/d$ increases
monotonically with $\lambda/d$, saturating at a constant value beyond $\lambda/d\sim
{\cal O} (10)$. When $\lambda/d\!\rightarrow\!\infty$, any significant variation in the
velocity and pressure fields will be localized near the plane $x=(\lambda \!-\! a)/2$,
where the BCs change from no-shear to finite slip. Since $p(x\!=\!0)=p(x\!=\!\lambda/2)$,
the longitudinal average of the lateral pressure gradient within the cell must vanish
(i.e. $\langle p_x\rangle\!=\!0$) and any pressure gradient above the surface of no-shear
will be canceled by an opposing gradient above the surface of finite slip. Since the
transition region in the vicinity of the the plane $x=(\lambda \!-\! a)/2$ does not
contribute significantly in the limit $\lambda/d\!\rightarrow\!\infty$, the condition
$\langle p_x\rangle\!=\!0$ is equivalent to the condition
\begin{equation}
\frac{\lambda-a}{2} (p_x)_1 = - \frac{a}{2} (p_x)_2~, \label{balance}
\end{equation}
where the subscripts 1 and 2 refer to the regions above the surface of no-shear (1) and
finite slip (2).

Now we first consider the case $b/d\!=\!0$. Since the flux must remain constant,
\begin{equation}\label{flux}
\int_0^d u_1 dz = \int_0^d u_2 dz~,
\end{equation}
where
\begin{eqnarray}
u_1 &=& U+\frac{(p_x)_1}{2\mu}~(z^2-d^2) \ \ \ \ \ \ \mbox{and}  \label{u1asympt}\\
u_2 &=& U\frac{z}{d}+\frac{(p_x)_2}{2\mu}~z(z-d)~. \label{u2asympt}
\end{eqnarray}
It follows that $(p_x)_2= 4 (p_x)_1 - 6 \mu U/d^2$, which when coupled with
Eqs.~(\ref{slipdefine}) and (\ref{balance}), yields the limiting value
\begin{equation}\label{across_limit_large_lambda1}
\lim_{\substack{b/d=0 \\ \lambda/d\!\rightarrow\!\infty}} \frac{L_s}{d} =
\frac{\left\langle u(z\!=\!0)\right\rangle }{ U-\left\langle u(z\!=\!0)\right\rangle} =
\frac{\lambda -a}{4a}\ .
\end{equation}
The same analysis can be extended to the case $b/d\!\neq\!0$ with the general result:
\begin{equation}
\lim_{\lambda/d\!\rightarrow\!\infty} \frac{L_s}{d}=\frac{\lambda d^2 + 8\lambda b d + 12
\lambda b^2 -ad^2 - 4abd}{4ad(d+3b)}~.\label{gentranslimit}
\end{equation}
The horizontal asymptotes (dotted lines) shown in Fig.~\ref{Couette_across_results}(a)
for $\lambda/d \!>\! 10$ represent solutions to Eq.~(\ref{gentranslimit}) for the
designated values of $b/d$ and $a/\lambda$.
%

In the opposite limit $\lambda/d\!\rightarrow\!0$, i.e. where the upper and lower walls
are essentially infinitely far apart, the deviation of the flow field from pure shear
flow over a {\em homogeneous} surface with slip length $L_s$ is limited to a thin layer
whose thickness scales with $\lambda$. As a consequence, the effective slip length should
be independent of the cell depth $d$ and independent of the particular mechanism used to
generate the flow, i.e. the same slip length should result for pressure-driven or
shear-driven flow. Lauga and Stone~\cite{Stone03} determined the asymptotic behavior of
the effective slip length for pressure-driven flow in a cylindrical tube of radius $R$
with periodically distributed (transverse) rings denoting alternating surfaces of
no-shear or no-slip ($b=0$):
\begin{equation} \label{slip_across_limit2}
\lim_{\substack{b/R=0 \\ \lambda/d \rightarrow 0}} \frac{L_s}{R} = \frac{\lambda}{2\pi
R}\, \ln\!\left( \frac{1}{\cos\left(
    \frac{\pi}{2} \frac{\lambda-a}{\lambda}\right)}\right)~.
\end{equation}
The solutions to Eq.~(\ref{slip_across_limit2}), obtained by replacing the capillary
radius $R$ with the cell depth $d$, superimpose perfectly (sloped dashed lines) onto the
full numerical solutions shown in Fig.~\ref{Couette_across_results}(a). In this limit,
the slip length increases linearly with $\lambda/d$ up to a limit $\lambda/d \approx 1$.

The effective slip length in the limit $\lambda/d\!\rightarrow\!0$ for the case $b
\!\neq\! 0$ can be derived as follows. When the array period $\lambda$ is much smaller
than the local slip length $b$, the slip velocity $u(x,z\!=\!0)$ should saturate towards
a constant value, $u_{s0}$, over the entire interval $0 \leq x \leq  \lambda/2$. Since
$u(z\!=\!0)\!=\!b u_z(z\!=\!0)$, it is also expected that the velocity gradient,
$u_z(z\!=\!0)$, will assume a constant value in the region $[\lambda - a)/2 \leq x \leq
\lambda/2, z=0]$. Consequently,
\begin{equation}\label{slip_across_limit3}
\frac{\partial \langle u \rangle}{\partial z}(0) = \frac{u_{s0}}{L_s} =
\frac{\lambda\!-\!a}{\lambda}\,0\,+ \frac{a}{\lambda} \frac{u_{s0}}{b}\ \ \ \Rightarrow\
\ \ \frac{L_s}{d} = \frac{\lambda}{a} \frac{b}{d}\, ,
\end{equation}
i.e. the effective slip length becomes independent of $\lambda/d$ for fixed $a/\lambda$.
The term proportional to $(\lambda\!-\!a)/\lambda$ accounts for the vanishing
contribution of the no-shear regions to $\partial \langle u \rangle/\partial z(z\!=\!0)$.
The horizontal dashed lines shown in Fig.~\ref{Couette_across_results}(a) for $\lambda/d
\!<\! 0.1$ represent solutions to Eq.~(\ref{slip_across_limit3}) for the designated
values of $b/d$ and $a/\lambda$.

In Fig.~\ref{Couette_across_results}(b) is plotted the effective slip length, $L_s/d$,
versus $a/\lambda$ for $b/d\!=\!0$ and $\lambda/d \!=\! 1.0$ and $14.3$. The data points
for $\lambda/d\!=\!14.3$ are scaled by a factor 0.3 for convenience. The effective slip
vanishes as $a/\lambda\!\rightarrow\!1$ since the surface coverage by regions of perfect
slip decreases to zero. The numerical results were compared to a Taylor expansion of
Eq.~(\ref{slip_across_limit2}) in the limit of $a/\lambda \!\rightarrow\! 0$
\begin{equation}\label{Lauga_analyt_along2}
\frac{L_s}{d} = - \frac{\lambda}{2\pi d} \left[ \ln\!\left( \frac{a}{\lambda}\right) +
\ln\!\left( \frac{\pi}{2}\right) \right].
\end{equation}
The dashed line shown in Fig.~\ref{Couette_across_results}(b) for $\lambda/d\!=\!1.0$
perfectly superimposes on the results of the full numerical solutions. The numerical
solution for $\lambda/d\!=\!14.3$ can also be approximated by a fit-function 
$A\log[a/\lambda]+B$ for $a/\lambda \ll 1$, as shown by the dashed line; however,
Eq.~(\ref{Lauga_analyt_along2}) no longer holds because $\lambda/d \nleq 1$.

Streamlines of the flow field, corresponding to the contour lines (i.e. constant values)
of the streamfunction, are shown in Figs.~\ref{Couette_across_results}(c,d). The left and
right panels represent the solutions for $b/d\!=\!0.048$ and $a/\lambda\!=\!0.5$ for (c)
$\lambda/d\!=\!1$ and (d) 20. The vertical line denotes the transition in boundary
condition at the lower wall from no-shear (left) to finite slip (right). For small values
of $\lambda/d \!\leq\! 1.0$, the streamlines are essentially horizontal in the larger
portion of the cell and the deviation of the streamfunction from pure Couette flow over a
homogeneous surface is confined to a small distance from the patterned wall. As
$\lambda/d$ increases, the perturbation extends further away from the lower boundary. For
large $\lambda/d$ the streamlines are horizontal above the individual stripes except for
a step-like vertical displacement in the vicinity of the transition point
$x=(\lambda\!-\!a)/2$.

\subsection{Longitudinal configuration} \label{Couette_long}

The velocity field corresponding to the longitudinal configuration shown in
Fig.~\ref{Couette_geometry}(b) is unidirectional and given by ${\bf u}(x,z)\!=\!(0,v,0)$.
There is no pressure gradient in this configuration and the numerical solutions are
derived directly from the Stokes equation $\nabla^2v\!=\!0$. The computational cell is
shown in Fig.~\ref{Fig_BCs}(b), where the direction of motion of the upper wall is
indicated by the white concentric circles. Only four BCs are required for solution of the
velocity field $v$. Aside from the obvious constraints of finite slip, $v_x$ must vanish
at $x\!=\!0$ and $x\!=\!\lambda/2$ because these are planes of mirror symmetry. The
complete set of BCs is given by:
\begin{eqnarray}
   v\left(\mbox{$0 \!\le\! x \!\le\! \frac{\lambda}{2}$},  z\!=\!d\right) & = & U \!-\! b v_z (x,z\!=\!d) \\
v_z \left(\mbox{$0 \!\le\! x \!\le\! \frac{\lambda-a}{2}$},z\!=\!0\right) & = & 0 \\
v\left(\mbox{$\frac{\lambda-a}{2} \!\le\! x \!\le\! \frac{\lambda}{2}$},z\!=\!0\right) & = & b v_z (x,z\!=\!0) \\
v_x (x\!=\!0,z) & = & 0 \ =\ v_x \left(\mbox{$\frac{\lambda}{2}$},z\right)~.
\label{sideBC}
\end{eqnarray}
Eq.~(\ref{sideBC}) and a lateral average of the Stokes equation across the computational
cell i.e. $\langle v(z) \rangle =(2/\lambda) \int_0^{\lambda/2} v(x,z)\, dx$, leads to
$\langle v \rangle_{zz}\!=\!0$. As in the transverse case, the averaged velocity profile,
$\langle v \rangle$, is therefore a linear function of $z$. Geometric similarity
determines the equation for the effective slip length, namely
\begin{equation} \label{slipdefinev}
\frac{L_s}{d} = \frac{\langle v(z\!=\!0) \rangle}{\langle v(z\!=\!d) \rangle - \langle
v(z\!=\!0) \rangle}~.
\end{equation}
%
\begin{figure}[t]
\begin{center}
\includegraphics[width=7.5cm]{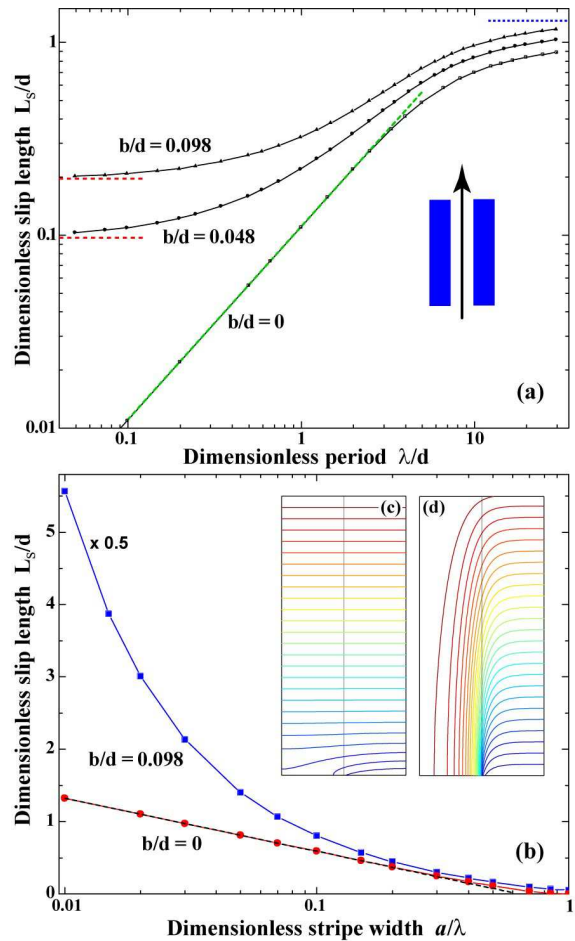}
\end{center}
\caption{(Color online) (a) Normalized slip length $L_s/d$ versus normalized pattern
period $\lambda/d$ derived from the Stokes solutions for the longitudinal flow
orientation. The parameter values shown are $b/d\!=\!0$, 0.048 and 0.098 and
$a/\lambda\!=\!0.5$. The straight line superimposed on the data for $b/d\!=\!0$
corresponds to Eq.~(\ref{Philip_analyt_along1}) where $L_s/d \!\sim\! \lambda/d$.
(b) Normalized slip length $L_s/d$ versus normalized stripe width $a/\lambda$ for
$\lambda/d\!=\!1$ and $b/d\!=\!0$ and 0.098. The straight line superimposed on the
numerical solutions corresponds to Eq.~(\ref{Philip_analyt_along2}). The data points for
$b/d\!=\!0.098$ are scaled by a factor 0.5 for convenience.
(c-d) Velocity contours corresponding to the longitudinal Stokes flow solutions for (c)
$\lambda/d\!=\!0.35$ and (d) 10, where $b/d\!=\!0.048$ and $a/\lambda\!=\!0.5$. The
domain size is $\lambda/2 \times d$; the thin vertical lines designate the boundary
between surfaces of no-shear (left) and finite slip (right).}
\label{Couette_along_results}
\end{figure}
Figure~\ref{Couette_along_results}(a) represents numerical results for the normalized
effective slip length, $L_s/d$, as a function of $\lambda/d$. Over the range shown,
$L_s/d$ increases monotonically with $\lambda/d$. As with the transverse geometry, there
is no significant increase in slip length beyond $\lambda/d\sim {\cal O} (10)$. The
absolute values of $L_s/d$ are larger than in the transverse case. This is due to the
fact that for unidirectional flow, the liquid above the region of no-shear always remains
in line with the frictionless stripes and is never subject to any deceleration caused by
the regions of finite slip. The functional dependence of $L_s/d$ on $\lambda/d$, however,
is identical to the transverse orientation. As $\lambda/d \!\rightarrow\! 0$, the slip
length should be independent of the cell depth, $d$, and independent of the type of flow
(whether pressure- or shear-driven). Using the analytical solutions of
Philip~\cite{Philip1,Philip2} for the longitudinal configuration, Lauga and
Stone~\cite{Stone03} extracted the effective slip length for pressure driven flow in a
cylindrical tube of radius $R$ in the presence of alternating stripes of no-shear and
no-slip ($b\!=\!0$):
\begin{equation}\label{Philip_analyt_along1}
\lim_{\substack{b/R=0 \\ \lambda/R \rightarrow 0}} \frac{L_s}{R} = \frac{\lambda}{\pi R}
\ln\!\left( \frac{1}{\cos\left(
    \frac{\lambda-a}{\lambda}\frac{\pi}{2}\right)}\right)~.
\end{equation}
The solutions to Eq.~(\ref{Philip_analyt_along1}), obtained by replacing the capillary
radius, $R$, with the planar cell depth, $d$, are almost indistinguishable from the
results of the full numerical solutions in Fig.~\ref{Couette_along_results}(a). In this
limit, the slip length $L_s/d$ is exactly twice that of the transverse configuration [see
Eq.~(\ref{slip_across_limit2})] and scales linearly with $\lambda/d$.

An analytic expression for the effective slip, $L_s/d$, can be derived in the limit
$\lambda/d\!\rightarrow\!\infty$, by examining the flow field above the patterned
substrates. The velocity profile above the no-shear surface (1) is plug-like and given by
$v_1(z)=U$. Above the surface of finite slip, $v_2(z) = U(z+b)/(2b+d)$. The latter
result is obtained by noting that the shear rate, $u_z$, is constant throughout the gap
depth and equal to $U/(2b+d)$. Calculating the average flow speed, $\langle v \rangle$,
at the upper and lower boundaries and substituting these into Eq.~(\ref{slipdefinev})
leads to the expression:
\begin{equation} \label{along_limit_large_lambda}
\lim_{\lambda/d \rightarrow \infty} \frac{L_s}{d} = \frac{\lambda -a}{a}+\frac{2\lambda
-a}{a}\, \frac{b}{d}~.
\end{equation}
The horizontal dashed line for $b/d=0.098$ and $\lambda/d \gtrsim 10$ corresponds to
Eq.~(\ref{along_limit_large_lambda}).

In Fig.~\ref{Couette_along_results}(b) is plotted the numerical solutions for $L_s/d$
versus  $a/\lambda$ for $\lambda/d\!=\!1$ and $b/d=0~\mbox{and}~0.098$. The data points
for $b/d=0.098$ have been scaled by 0.5 for convenience. For $b/d\!=\!0$ and small values
$a/\lambda$, a Taylor expansion of Eq.~(\ref{Philip_analyt_along1}) gives
\begin{equation} \label{Philip_analyt_along2}
\lim_{\substack{b/d = 0 \\ \lambda/d \rightarrow 0}} \frac{L_s}{d} = -\frac{\lambda}{\pi
d} \left[ \ln\!\left( \frac{a}{\lambda}\right) + \ln\!\left(\frac{\pi}{2}\right) \right].
\end{equation}
The straight line superimposed on the data in Fig.~\ref{Couette_along_results}(b)
represents the asymptotic values given by Eq.~(\ref{Philip_analyt_along2}). The agreement
with the analytical limit for $a/\lambda\!\lesssim \!0.3$ is very good.

Velocity contours, corresponding to constant values of $v$, are shown in
Figs.~\ref{Couette_along_results}(c,d). The left and right panels represent solutions for
(c) $\lambda/d\!=\!0.35$ and (d) $\lambda/d\!=\!10$ where $b/d\!=\!0.048$ and
$a/\lambda\!=\!0.5$. The vertical line denotes the position corresponding to the change
in boundary condition at the lower wall from no-shear (left) to finite slip (right). For
$\lambda/d\!\leq \!0.35$, the velocity contours are horizontal throughout almost the
entire cell and the deviations from pure shear flow over a homogeneous surface are
confined to a small distance from the patterned wall. For $\lambda/d\!=\!10$, the
perturbation extends vertically across the cell. For large $\lambda/d$, the velocity
distribution varies from plug--like above the region of perfect slip to Couette--like
above the region of finite slip, as assumed in the derivation leading to
Eq.~(\ref{along_limit_large_lambda}) for $\lambda/d \rightarrow \infty$.

\section{MD simulations and parameter values} \label{sec:MDDetails}

We have previously used MD simulations to investigate what equilibrium parameters control
the degree of slip in simple and polymeric fluids and how the slip length depends on
shear rate~\cite{Nature97,Priezjev04}. In these previous studies, the wall--fluid
potential was spatially homogeneous. In this current work, we examine the behavior of the
effective slip length for a fluid subject to planar shear in the presence of a
heterogeneous bottom wall for the two flow configurations shown in
Fig.~\ref{Couette_geometry}. The wall--fluid interactions are adjusted to mimic
alternating stripes of finite slip and no--shear by adjusting the attractive part of the
potential to simulate more attractive and less attractive regions. The MD simulations
described next were conducted with the LAMMPS numerical code~\cite{Lammps}. In what
follows, we refer to the more attractive surface as {\it wetting} and the less attractive
surface as {\it non}--wetting.

\begin{figure}[b]
\includegraphics[width=8.0cm,height=6.0cm]{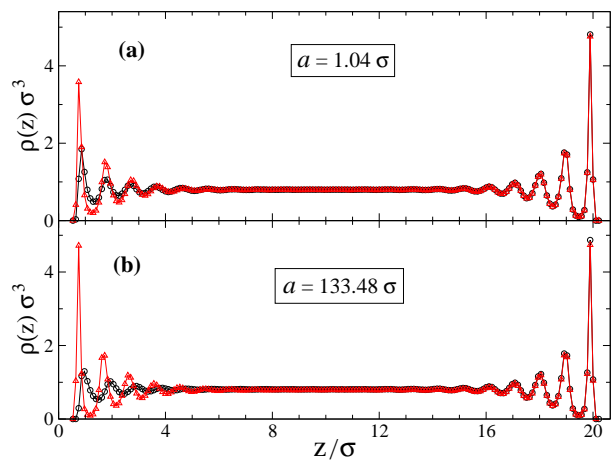}
\caption{(Color online) Average normalized fluid density, $\rho(z) \sigma^3$, above the
wetting ($\delta_{\rm wf}\!\!=\!1.0$: triangles) and non--wetting stripes ($\delta_{\rm
wf}\!\!=\!0.1$: circles). The parameter values shown are (a) $a\!=\!1.04\,\sigma$ and (b)
$a\!=\!133.48\,\sigma$ for $\varepsilon_{\rm wf}/\varepsilon\!=\!0.8$.} \label{MD_dens}
\end{figure}

The simulation cell consisted of $30720$ fluid molecules interacting through a
Lennard--Jones (LJ) potential,
\begin{equation}
V_{LJ}(r)=4\varepsilon \bigg[\bigg(\frac{\sigma}{r}\bigg)^{12}-
\delta~\bigg(\frac{\sigma}{r}\bigg)^{6}~\bigg]~, \label{LJ}
\end{equation}
where $\varepsilon$ and $\sigma$ represent the energy and length scales characteristic of
the fluid phase. The cut-off radius was set to $r_c\!=\!2.5\,\sigma$. The parameter
$\delta$, which controls the attractive part of the potential for fluid--fluid
interactions, was held fixed at $\delta\!=\!1$. The wall--fluid (wf) parameters were
chosen to be $\sigma_{\rm wf}\!=\!0.75\, \sigma$ and $\varepsilon_{\rm
wf}/\varepsilon\!=\!0.8$, 0.9 or 1.0. Surfaces of finite slip in the hydrodynamic
analysis corresponded to the parameter value $\delta_{\rm wf}\!=\!1.0$ (i.e. wetting);
surfaces of no-shear (or likewise perfect slip) corresponded to the value $\delta_{\rm
wf}\!=\!0.1$ (i.e. non-wetting). For the MD simulations, we restricted our study to the
case $a/\lambda\!=\!1/2$ such that the wetting and non-wetting portions of the substrate
occupy equal areas.

\begin{figure}[t]
\includegraphics[width=8.0cm,height=6.0cm]{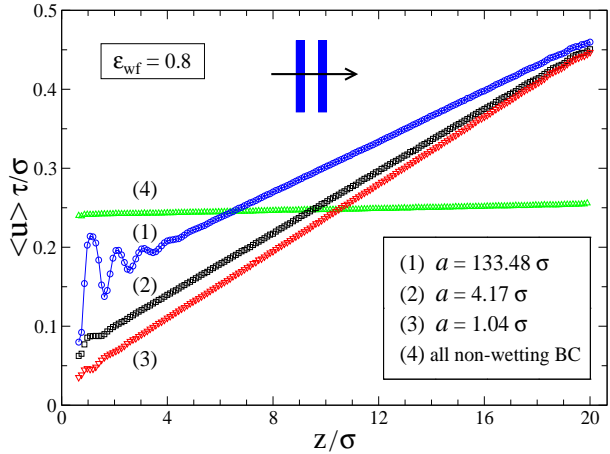}
\caption{(Color online) Average normalized velocity profile, $\langle u \rangle\,
\tau/\sigma$ corresponding to the transverse flow configuration for $\varepsilon_{\rm
wf}/\varepsilon\!=\!0.8$ and $a/\sigma = 1.04$, 4.17 and 133.48. The nearly horizontal
profile shown in (4), which designates a plug-like profile representative of surfaces
with large slip lengths, was obtained by setting the wall-fluid potential parameter to
$\delta_{\rm wf}=0.1$ along both walls. This choice effectively reproduces non-wetting
bounding walls.} \label{MD_velo_trans}
\end{figure}

The upper and lower walls of the simulation cell each consisted of $12288$ molecules
distributed between two (111) planes of an FCC lattice with density $\rho_{w}\!=\!4\rho$,
where $\rho\!=\!0.81\,\sigma^{-3}$ is the density of the fluid phase. The fluid was
confined to a fixed height $d=20.15\,\sigma$; the cell volume was $266.96\,\sigma\times
7.22\,\sigma \times d$ for the transverse geometry. To eliminate any finite size effects
for the longitudinal geometry, the system size along the $\hat{y}$-axis was doubled in
length to $14.45\,\sigma$, requiring simulations with $61440$ fluid molecules. For either
configuration, periodic BCs were enforced along the $\hat{x}$ and $\hat{y}$ axes. The
fluid was held at a constant temperature $T\!=\!1.1\,\varepsilon/k_B$ by means of a
Langevin thermostat \cite{Grest86} with a friction coefficient $\tau^{-1}$. Here, $k_B$
is the Boltzmann constant. This damping term is only applied to the coordinate equation
perpendicular to the direction of flow~\cite{Thompson90,Nature97}. The equations of
motion were integrated using the Verlet algorithm~\cite{Allen87} with a time step
$\triangle t\!=\!0.005\,\tau$, where $\tau\!=\!\sqrt{m\sigma^2/\varepsilon}$ represents
the characteristic time set by the LJ potential and $m$ is the monomer mass.
The fluid was subject to steady planar shear by translating the upper wall at a constant
speed $U$; the lower, patterned wall remained stationary. In all the simulations, the
speed of the upper wall was held fixed at $U=0.5\, \sigma/\tau$. After an equilibration
period exceeding $10^4\tau$, the fluid velocity profile was obtained by averaging the
instantaneous monomer speeds in slices $\Delta z= 0.1\,\sigma$ for a time interval
$\Delta t \approx 3\cdot10^4\tau$. The Reynolds number, based on the upper wall speed
$U$, the wall separation $d$ and the fluid shear viscosity (determined previously
\cite{Nature97,Priezjev04} to be $\mu = 2.2 \pm 0.2 \epsilon \tau / \sigma^3$ for
comparable shear rates) was estimated to range from $2\!-\!5$, indicative of negligible
inertial effects and laminar flow conditions. In fact, this estimate provides only an
upper bound on the Reynolds number, since the actual fluid velocity for surfaces
comprising regions of finite and infinite slip is significantly smaller than the upper
wall speed. In our studies, use of the fluid flow speed further reduces ${\sf Re}$ by a
factor of up to 2. We conclude that the small Reynolds numbers characterizing the MD
simulations are consistent with the theoretical restriction for the Stokes flow solutions
obtained in the limit ${\sf Re}\!=\!0$. We also note that the numerical solutions to the
Stokes equation for the longitudinal geometry are valid irrespective of the value of the
Reynolds number because the unidirectional flow causes the inertial term in the Navier-Stokes
equation to vanish identically.

\begin{figure}[t]
\includegraphics[width=8.0cm,height=6.0cm]{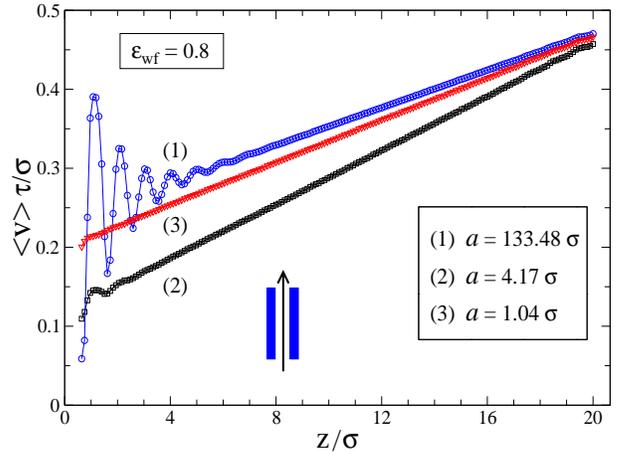}
\caption{(Color online) Average normalized velocity profile, $\langle v \rangle\,
\tau/\sigma$, corresponding to the longitudinal flow configuration for $\varepsilon_{\rm
wf}/\varepsilon\!=\!0.8$ and $a/\sigma = 1.04$, 4.17 and 133.48.} \label{MD_velo_para}
\end{figure}

\section{Results of MD simulations for Transverse and Longitudinal Flow}
\label{sec:MDResults}

The two sets of curves in Fig.~\ref{MD_dens} show the average normalized fluid density,
$\rho(z) \sigma^3$, for the transverse flow configuration in the region above the wetting
and non--wetting stripes for $\varepsilon_{\rm wf}/\varepsilon\!=\!0.8$ and $a/\sigma =
1.04~\mbox{and}~133.48$. The choice $a\!=\!133.48\,\sigma$ represents the accommodation
of only two stripes at the lower wall within the Couette cell. The oscillations near the
upper and lower boundaries reflect the molecular layering caused by the presence of dense
walls ~\cite{Thompson90}. Increasing the attractive part of the LJ potential generates
larger peak maxima and more oscillations. Above either type surface, the density
oscillations persist for about $4-6$ molecular diameters from the wall. Decreasing the
strength of the attractive interaction shifts the first peak maximum away from the lower
wall. Also, the fluid density above the wetting stripes is found to increase with
$a/\sigma$. The density profiles corresponding to longitudinal flow configurations are
quite similar to the ones shown here.

Figure~\ref{MD_velo_trans} shows representative velocity profiles across the cell depth
for transverse flow with $\varepsilon_{\rm wf}/\varepsilon\!=\!0.8$ and
$a/\sigma=1.04,~4.17~\mbox{and}~133.48$. Shown for comparison is the velocity profile
corresponding to the case of {\it uniformly} non-wetting walls where $\delta_{\rm
wf}=0.1$ holds for both surfaces. Decreasing the wall--fluid interaction leads to a high
degree of slip and a plug--like velocity field. The remaining three profiles increase
linearly with $z/\sigma$, as expected for a fluid subject to planar shear, except in the
vicinity of the lower wall. Significant deviations from linearity occur for large stripe
widths. These oscillations are caused by the difference in the positions of the fluid
density maxima above the wetting and non--wetting regions [see Fig.~\ref{MD_dens}(b)]. As
evident from the velocity profile, the degree of slip increases with increasing values of
$a$.

\begin{figure}[t]
\includegraphics[width=8.0cm,height=6.0cm]{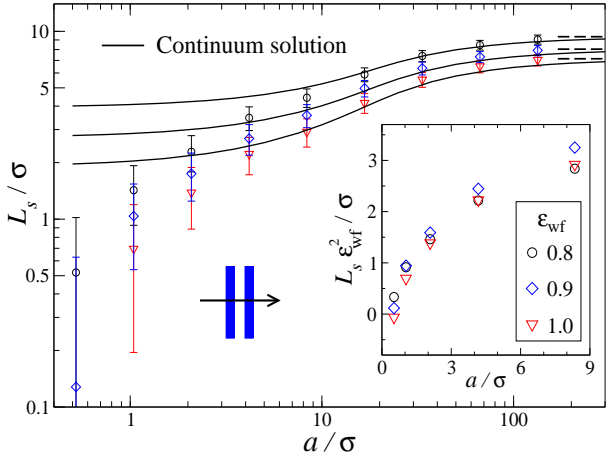}
\caption{(Color online) Comparison of the effective slip length, $L_s$, as extracted from
the MD simulations (symbols), with numerical solutions of the Stokes equation
(continuous lines) for transverse flow. The local slip length, $b$, as extracted from the
MD simulations, decreases with increasing wall--fluid attraction energy, namely $b/\sigma
\!=\!1.97$, 1.36 and 0.95  for $\varepsilon_{\rm wf}/\varepsilon\!=\!0.8$ ($\circ$), 0.9
($\diamond$) and 1.0 ($\triangledown$), respectively. The dashed horizontal lines for
$a/\sigma \gtrsim 100$ correspond to Eq.~(\ref{gentranslimit}). Inset: MD results showing
collapse of the effective slip length $L_s/\sigma$ when rescaled by the quantity
${\varepsilon}^2_{\rm wf}$, versus $a/\sigma$.} \label{hydro_Navier_trans}
\end{figure}

Figure~\ref{MD_velo_para} shows the computed velocity profiles for longitudinal
configurations. The behavior is similar to that shown in Fig.~\ref{MD_velo_trans} for the
transverse orientation, but the amplitude of the oscillations near the lower wall is
significantly larger. In this case, the degree of slip does not increase monotonically
with $a$. The smallest stripe width generates the second largest slip velocity in
Fig.~\ref{MD_velo_para}. As the stripe width increases, it is found that the wetting
regions induce stronger molecular ordering in the first fluid layer adjacent to the wall,
causing a reduction in the slip length, as noted in Fig.~\ref{structure}.

For direct comparison to the hydrodynamic predictions, it was necessary to extract the
actual values of the local slip length, $b$, representative of the surfaces characterized
by $\delta_{\rm wf}=1.0$, for input values to the boundary conditions used in computing
the solutions to the Stokes equation. This was accomplished in the MD simulations by
extrapolating the average velocity profile at the \textit{top} wall to a speed $U$ for
different values of $a$ imposed on the lower wall. The extrapolated distance $b$ was
found to depend on the wall--fluid interaction energy but not the shear rate in the fluid
nor the flow orientation. As expected, the values of $b$ decreased with increasing value
of the wall--fluid interaction energy, namely $b/\sigma \!=\!1.97$, 1.36 and 0.95  for
$\varepsilon_{\rm wf}/\varepsilon\!=\!0.8$, 0.9 and 1.0, respectively. By contrast, the
local slip length for the flat velocity profile shown in Fig.~\ref{MD_velo_trans} for
uniformly non-wetting walls was found to be $(362\pm 10)\,\sigma$. Given that this slip
length significantly exceeds the wall separation, the choice $\delta_{\rm wf}\!=\!0.1$
approximates very well the behavior of surfaces of perfect slip (i.e. no--shear) assumed
in the continuum calculations.

\begin{figure} [t]
\includegraphics[width=8.0cm,height=6.0cm]{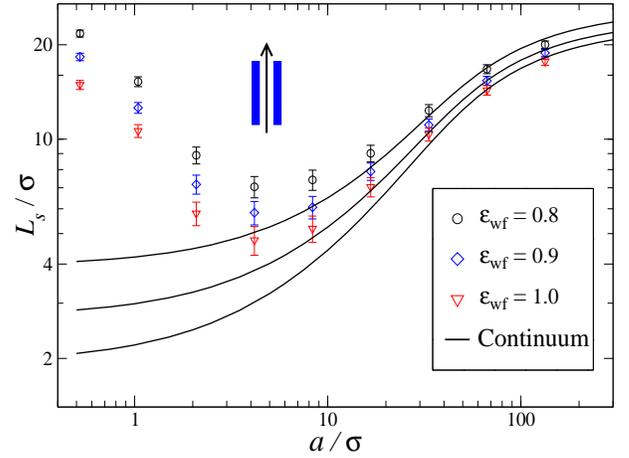}
\caption{(Color online) Direct comparison of the effective slip length extracted from the
MD simulations (symbols) with the numerical solutions of the Stokes equation (continuous
lines) for longitudinal flow. The local slip length, $b$, as extracted from the MD
simulations, varies with the LJ wall--fluid interaction energy, $\varepsilon_{\rm wf}$ as
$b/\sigma \!=\!1.97$, 1.36 and 0.95  for $\varepsilon_{\rm wf}/\varepsilon\!=\!0.8$
($\circ$), 0.9 ($\diamond$) and 1.0 ($\triangledown$), respectively. The local slip
lengths are observed to be independent of the flow orientation.}
\label{hydro_Navier_para}
\end{figure}

The composite or effective slip length, $L_s$, was determined in the MD simulations by
linear extrapolation below the stationary lower surface of the velocity profile to the value zero.
Figure~\ref{hydro_Navier_trans} represents a plot of $L_s/\sigma$ with increasing
normalized stripe width, $a/\sigma$, and increasing wall--fluid interaction strength,
$\varepsilon_{\rm wf}$, for transverse flow configurations. The MD results (symbols) show
a sharp increase in slip length for $a/\sigma \lesssim 10$ and saturation to a constant
value beyond $a/\sigma \gtrsim 100$.

\section{Discussion}

As described in Section~\ref{Couette_trans} and for fixed values of $a/\lambda$, the
effective slip length derived from hydrodynamic considerations depends only on the ratios
$\lambda/d$ and $b/d$. The molecular length scale, $\sigma$, plays no part in the
analysis. For direct comparison to the MD results, it was therefore necessary to multiply
the numerical values of $L_s/d$, $\lambda/d$ and $b/d$ from the Stokes solutions with the
value of the wall separation, $d\!=\!20.15\,\sigma$, used in the MD simulations. The
largest ratio, $a/d\!=\!6.62$, accessible to the MD simulations was only limited by
computational resources. The solid lines shown in Fig.~\ref{hydro_Navier_trans} represent
solutions of the Stokes flow equation for transverse flow, as discussed in
Section~\ref{Couette_trans}. The agreement between the continuum predictions and the MD
simulations is excellent for $a/\sigma \gtrsim {\cal O}(10)$; significant deviations
occur for $a/\sigma \lesssim {\cal O}(1)$. The asymptotic predictions given by
Eq.~(\ref{gentranslimit}) for $\lambda/d \!=\! 2a/d \!\rightarrow\! \infty$ are
designated by the dashed horizontal lines in Fig.~\ref{hydro_Navier_trans}.

The Green-Kubo type analysis of Barrat and Bocquet~\cite{Barrat94,Barrat99} for
homogeneous surfaces characterized by a single wall--fluid interaction energy predicts
that the slip length scales as $\varepsilon_{\rm wf}^{-2}$ provided the in--plane
structure factor, fluid contact density, and in--plane diffusion coefficient
characteristic of the first fluid layer remain relatively constant. The results shown in
the inset of Fig.~\ref{hydro_Navier_trans} for the transverse geometry confirm this
prediction for the range $a\lesssim 10\,\sigma$, even for the case of a composite
potential where the wall--fluid interaction alternates between two values of $\delta_{\rm
wf}$. This collapse fails above $a\!\gtrsim\! 10\,\sigma$ where the continuum solutions
show excellent agreement with the molecular simulations. This behavior suggests that for
$a/\sigma \!\lesssim\! {\cal O}(10)$, the effective slip length is mostly determined by
the molecular scale frictional properties between the first fluid layer and the lower
wall. For $a/\sigma \!\gtrsim\! {\cal O}(10)$, however, the effective slip length is set
by the wall separation $d$, the pattern lengthscales $a$ and $\lambda$ and the local slip
length $b$. The transition region $8\lesssim a/\sigma \lesssim 30$ therefore contains
mesoscopic information from both the molecular and hydrodynamic descriptions.

The deviation between the MD simulations and the Stokes solutions below $a/\sigma
\!\lesssim\! {\cal O}(10)$ can be understood as follows. The lower wall is comprised of a
potential whose interaction strength alternates between wetting and non--wetting values
with a periodicity set by the stripe width $a$, which approaches the molecular scale. The
fluid molecules no longer experience uninterrupted stretches of wetting and non-wetting
regions; instead, the fluid molecules are exposed to an effectively roughened surface
with molecular scale corrugations. These corrugations trap the fluid molecules, thereby
suppressing slip at the wall--fluid interface. The commensurability between the fluid
molecular size and the wall corrugation size can in fact lead to a no-slip condition for
slightly larger values of $\varepsilon_{\rm wf}$~\cite{Priezjev05}. It is therefore not
surprising that the effective slip length for the transverse configuration, as shown in
Fig.~\ref{hydro_Navier_trans}, decreases sharply with decreasing values of $a$. This
effect also explains why for the smallest values of $a$ the slip length $L_s$ is even 
smaller smaller than the local slip length obtained for a fluid confined between two 
identical walls both characterized by the same value $\delta_{\rm wf}=1.0$. For example, for $\varepsilon_{\rm wf}/\varepsilon=0.8$ and $a/\sigma\!=\!1.04$, we find that 
$b/\sigma\!=\!1.97$ but $L_s/\sigma\!=\!0.5$!
\begin{figure}[b]
\includegraphics[width=8.0cm,height=6.0cm]{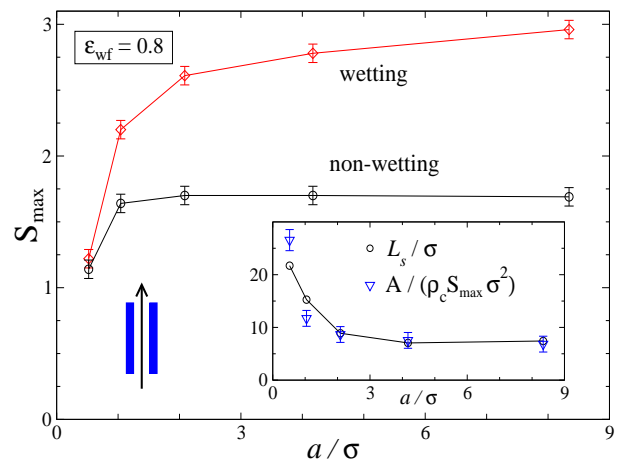}
\caption{(Color online) The dominant peak in the in--plane fluid structure factor
evaluated separately above the wetting ($\diamond$) and non--wetting ($\circ$) regions
for longitudinal flow and $\varepsilon_{\rm wf}/\varepsilon\!=\!0.8$. Inset: MD results
showing the strong correlation between the effective slip length $L_s/\sigma$ (circles:
data from Fig.~\ref{hydro_Navier_para}) and the quantity $A(S_{\max}\rho_c
\sigma^2)^{-1}$ (triangles), which characterizes the degree of molecular ordering within
the first fluid layer above the wetting stripes. The value of the fitting parameter $A$
is 92.4.} 
\label{structure}
\end{figure}

Figure~\ref{hydro_Navier_para} shows the behavior of $L_s/\sigma$ as a function of stripe
width, $a/\sigma$, and increasing interaction strength, $\varepsilon_{\rm wf}$, for
longitudinal flow configurations. The results of the MD simulations (symbols) show a
sharp decrease in slip length below $a/\sigma \!\sim\! 10$ followed by a rapid rise. The
effective slip lengths have similar magnitudes for very small and very large values of
$a$. Once again, there is excellent agreement between the Stokes flow solutions and the
MD simulations for $a/\sigma \!\gtrsim\! {\cal O}(10)$ but strong deviations below this
value. In contrast to the transverse configuration, however, the MD results predict much
larger effective slips than the continuum solutions for $a/\sigma \lesssim {\cal O}(10)$.
Because of the translational invariance of the flow inherent in this case, the molecular
scale roughness set by the composite potential at the bottom wall cannot diminish the
slip length. The reduction in molecular ordering above the wetting regions with
decreasing stripe width, however, leads to an increase in the slip length which exceeds
the slip lengths obtained for the transverse configuration as well as the continuum
predictions.

Previous MD simulations of Newtonian and non-Newtonian fluids have demonstrated that the
slip length for surfaces characterized by a single wall--fluid potential correlates
strongly with the degree of molecular ordering in the first fluid layer adjacent to the
wall \cite{Thompson90,Thompson92,Thompson95,Barrat94,Nature97,Barrat99,Priezjev04}. The
more orderly the molecular organization, as reflected by the maximum value of the
in-plane structure function, $S_{\max}$, the smaller the slip length. To test these
predictions for the case of patterned walls in a longitudinal orientation, we computed
the maximum value of the in-plane structure function within the first fluid layer above
the wetting and non-wetting regions separately. The thickness of the first fluid layer
was estimated from the position of the first minimum in the density profile above a
wetting stripe. The contact density $\rho_c$ was identified with the maximum of the fluid
density within the first fluid layer. The structure function was computed according to
$S(q)\!\!=\!|\sum_1^{N_{\ell}} e^{iqy}|^2/N_{\ell}$, where $N_{\ell}$ is the number of
molecules in the first fluid layer adjacent to either a wetting or non--wetting surface.
As shown in Fig.~\ref{structure}, the molecular ordering adjacent to a wetting region is
far stronger and increases with increasing stripe width, $a$. By contrast, the molecular
ordering adjacent to the non--wetting region is unaffected by the stripe width, $a$,
except for the smallest value shown. The inset in Fig.~\ref{structure} demonstrates the
correlation between the effective length and the parameter $(S_{\max}\rho_c
\sigma^2)^{-1}$ as estimated above the wetting regions. Here, the values of $L_s/\sigma$
versus $a/\sigma$ from Fig.~\ref{hydro_Navier_para} are plotted alongside the quantity
$A\,(S_{\max}\rho_c \sigma^2)^{-1}$, where $A\!=\!92.4$ is a fitting parameter. In the
limit $a/\sigma \!\lesssim\! {\cal O}(10)$, the strong correlation between $L_s$ and
$A\,(S_{\max}\rho_c \sigma^2)^{-1}$ establishes that the increase in effective slip
length for narrow stripe widths is mainly caused by the reduction in molecular ordering
within the first fluid layer above the wetting zones.

The BCs used in the continuum analysis correspond to stripes of finite (or no) slip and no shear (i.e. $b=\infty$). We repeated the analysis in Section II by replacing the no-shear BC with a second slip BC to define surfaces comprising alternating stripes of small ($b/\sigma = 1.97$) and large slip ($b/\sigma = 362$, as extracted from case (4) shown in Fig. 6). For the transverse configuration, the curve corresponding to $b/\sigma = 1.97$ in Fig. 8 showed a slight decrease in $L_s$ of about 3\% for $a/\sigma > 30$, whereas the longitudinal configuration generated a decrease of up to 9\% with respect to the values shown in Fig.~\ref{hydro_Navier_para}.

\section{Summary}

We have investigated the behavior of the slip length in Newtonian liquids subject to
planar shear in a Couette cell with mixed surface boundary conditions. The upper wall is
modelled as a homogenous surface with finite or no-slip moving at a constant speed above
a lower stationary wall patterned with alternating stripes representing regions of
no-shear and finite or no-slip. The velocity fields and effective slip lengths are
computed both from molecular dynamics (MD) simulations and solution of the Stokes
equation for flow parallel (longitudinal case) or perpendicular
(transverse case) to the stripe pattern. Detailed comparison between the results of
the hydrodynamic calculations and MD simulations shows excellent agreement when the
length scale of the substrate pattern geometry is larger than ${\cal O}(10\,
\sigma)$, where $\sigma$ denotes the fluid molecular diameter as set by the Lennard-Jones
interaction. The effective slip length then increases monotonically with $a/\sigma$ to a
saturation value. For the transverse case, the Stokes flow solutions predict an effective
slip larger than the MD results when $a/\sigma \sim {\cal O}(10)$. This discrepancy is
understood from a molecular point of view since a narrowing of the regions subject either
to no-shear or no-slip essentially establishes a roughened surface. The molecular scale
corrugation created by the composite wall potential strongly reduces the effective slip
length below the hydrodynamic results. This surface roughening effect is not present for
the longitudinal flow configuration since the fluid molecules are transported along
homogeneous stripes representing regions of either no-shear or finite slip. In this case,
however, the 2D fluid structure factor above the non-wetting stripes (regions of perfect
slip or equivalently no-shear) decreases for $a/\sigma \lesssim {\cal O}(10)$, which
enhances the effective slip lengths above the values predicted by the hydrodynamic
solutions. On the molecular level, the strong correlation observed between the
effective slip length and the product $(\rho_c S_{\rm max})^{-1}$ confirms that a reduction in
molecular ordering within the first fluid layer generates an increase in the effective
slip length.

Detailed comparison between continuum computations and molecular dynamics simulations is
of increasing importance to the development of hybrid computational
schemes~\cite{Thompson,Hadjiconstantinou,Feder,Nie04}. These algorithms are designed to
stitch together hydrodynamic solutions obtained from continuum equations with the
molecular scale solutions obtained from MD simulations or other microscopic solvers. It
has been demonstrated that the spatial coupling across this wide range in length scales
can be achieved by implementation of constraint dynamics within an overlap region. We
hope that our studies of shear driven flow along surfaces with mixed boundary conditions
will complement ongoing efforts using hybrid codes. The system and results described here
offer an interesting test case for better understanding of the intermediate region
bridging the behavior of fluids from the nanoscale to microscale dimensions.

\begin{acknowledgments}
The authors kindly acknowledge financial support from the National Science Foundation,
the NASA Microgravity Fluid Physics Program and the US Army TACOM ARDEC. NVP would like
to thank J.~Rottler for useful discussions. SMT gratefully acknowledges the support and
generous hospitality of the Moore Distinguished Scholar Program at the California
Institute of Technology.
\end{acknowledgments}

\end{document}